\documentclass[12pt]{article}
\usepackage[top = 2 cm, bottom = 2.5 cm, left = 2.5 cm, right = 1.5 cm]{geometry}
\usepackage{authblk, cite, color, amssymb, amsmath}
\usepackage[hypertex, colorlinks = true, linkcolor = blue, citecolor = red]{hyperref}
\usepackage[dvipsnames]{xcolor}

\begin{document}

\title{Can Quantum Particles Cross a Horizon?}

\author{\bf Merab Gogberashvili}
\affil{\small Javakhishvili Tbilisi State University, 3 Chavchavadze Avenue, Tbilisi 0179, Georgia \authorcr
Andronikashvili Institute of Physics, 6 Tamarashvili Street, Tbilisi 0177, Georgia}

\maketitle

\begin{abstract}

The prevalent opinion that infalling objects can freely cross a black hole horizon is based on the assumptions that the horizon region is governed by classical General Relativity and by specific singular coordinate transformations it is possible to remove divergences in the geodesic equations. However, the coordinate transformations usually used to demonstrate the geodesic completeness are of class $C^0$, while the standard causality theory requires that the metric tensor to be at least $C^1$. Introduction of $C^0$-class functions leads to the appearance of the additional delta-like sources in the Einstein equations and in the equations for quantum particles. Therefore, to explore the horizon region, in addition to the classical geodesic equations, one needs to use equation of quantum particles. Applying physical boundary conditions at the Schwarzschild and Kerr event horizons, we show existence of the exponentially decay/enhanced solutions (with the complex phases) to the Klein-Gordon equation. This means that in semi-classical approximation particles probably do not enter the black hole horizon, but are absorbed/reflected by it. Then it follows that the minimal classical size of any isolated body is its horizon radius, what potentially can solve main black hole mysteries.

\vskip 3mm
PACS numbers: 04.70.-s; 04.20.Dw; 04.62.+v
\vskip 1mm

Keywords: Black holes; Singularities; Fields in curved spacetime
\end{abstract}
\vskip 5mm


\section{Introduction}

Spacetime singularities are an inevitable feature for most of the solutions of the Einstein equations \cite{Haw-Pen}. The famous singularity theorems show geodesic incompleteness (currently the most acceptable criterion of existence of geometrical singularities) and assume that so called coordinate singularities (for which the curvature invariants are finite) can be avoided 'repairing' geodesics by specific singular coordinate transformations \cite{San-Gar, Haw-Ell}. At the same time singularity theorems say little about the nature of the singularity, in some cases the curvature blows up when differentiability dropping below the class $C^2$.

The most important singular solution to the Einstein equations is the spherically symmetric metric, which indicate the existence of Black Holes (BH). The classical definition of a BH as the complement of the causal past of future null infinity \cite{Haw-Ell} defines a BH spacetime rather than a BH as a normal source in some spacetime. However, the basic idea behind General Relativity is that geometry does not exist separately from matter and any physical solution of Einstein equations, including spherically symmetric metric, should be associated with some realistic matter source. True vacuum is an approximation, i.e. to vacuum spherically symmetric metric should correspond a point-like source and scalar curvature must be a $\delta$-function rather than zero as is mentioned in the literature \cite{delta-1, delta-2}. However, this assumption leads to some paradoxes, e.g. it was found that the spherically symmetric vacuum solutions of Einstein's equations is not diffeomorphic to the metric corresponding to the gravitational field of a $\delta$-like source \cite{Castro}.

In general, non-trivial choices of coordinates lead to different solutions of Einstein's equations. In so called Schwarzschild's gauge, the point-like source of gravity is located at the origin and its surface area is zero. While for other coordinates one would arrive at a rather grotesque picture where the point mass has a finite surface are! To avoid this problem some authors even claim that a neutral point-like source should have a zero gravitational mass \cite{ADM, Mitra-1, Mitra-2}. So in General Relativity there are problems with proper descriptions of coordinates and realistic sources of gravity, including BHs \cite{ADM, Mitra-1, Mitra-2, Gog-Mod}.

Recent imaging of the supermassive BH \cite{EHT} and the detection of gravitational waves from BH mergers \cite{LIGO}, have opened new observational windows to the regime of strong gravity and can give glue on properties of BHs. It has been widely believed that, due to the smallness of the curvature the horizon area of a large BH is governed by classical General Relativity. Based on the analysis of the classical geodesic equations it was found that the BH is characterized by a surrounding event horizon, which acts as a one-way membrane allowing particles in from the outside and preventing anything from escaping from the inside \cite{BH-1, BH-2, BH-3}. However, the assumption that particles can freely fell through the BH event horizon contradicts a unitary quantum theory and leads to the BH information paradox (see the recent review \cite{Mar}). The inconsistency of BH evolution with the principles of quantum mechanics strongly motivates the need to modify the classical description of BHs already at horizon scales \cite{Modif-1-1, Modif-1-2, Modif-2, Modif-3}. Spacetime singularities should be associated with reaching the limits of the physical validity of General Relativity, i.e. quantum effects can be expected to come in. It seems natural to replace any singular point in Riemann spacetime by a matter source, e.g. nobody worries about central singularity of the Coulomb potential in classical electromagnetism. So, there is a phenomenological possibility that quantum gravity effects replace the central singularity of spherically symmetric metric by an object of finite size. The question is what is the radius of this core?

From exploring the supermassive BH image and from numerical simulations of a BH merger \cite{merger} one can conclude that BHs can be considered as horizon size massive spheres. This horizon radius is what enters the expression of BHs moment of inertia \cite{Ha}, while in conventional treatment all mass is located at the center. Also, in standard approach neither event nor apparent horizons can actually send signals to infinity: apparent horizons lie inside event horizons which in turn are the boundary for signals that can reach infinity \cite{Haw-Ell}. It is not event horizons themselves that interact, since they are considered as empty space just screening the central sources. Since we assume that quantum effects are strong already at the BH horizon, it is insufficient to explore this region only with the classical geodesic (Hamilton-Jacobi) equations and one should use at least semi-classical approximation of the equations for quantum particles on curve background \cite{Gog-Pan, Gog}.

In this paper we discuss the prevalent opinion that interior matter is absent at the horizon of a BH and ingoing particles are able freely penetrate it. In Sec.~\ref{Sing-Coord} we recall that it is unacceptable to use $C^0$-class coordinate functions in classical General Relativity. In Sec.~\ref{AB} the analogy of the vacuum $C^0$-metrics with the topologically non-trivial gauge fields is indicated. In Sec.~\ref{Sin-Sch} it is shown that the Regge-Wheeler's 'tortoise' radius and the singular coordinate transformations, used to demonstrate continuity of geodesics at a BH horizon, are unacceptable class $C^0$ functions and lead to the appearance of fictitious sources in the Einstein equations. In Sec.~\ref{KG-HJ} we present the arguments that to explore the horizon region of a BH one needs to use solutions of the exact second order equation of quantum particles (e.g. Klein-Gordon equation) and not only the first order geodesic (Hamilton-Jacobi) equation. In Sec.~\ref{KG-Sch} and \ref{KG-Kerr} the solutions of the Klein-Gordon equation on the Schwarzschild and Kerr backgrounds, under physical boundary conditions, are presented. These solutions contain the real-valued exponentially time-dependent factors (with the complex phases), what means that in semi-classical approximation particles probably do not enter the BH event horizon, but are absorbed and some are reflected by it, what potentially can solve the main BH mysteries. The Sec.~\ref{Concl} is devoted to the concluding remarks.


\section{On the singular coordinate transformations} \label{Sing-Coord}

Recall that a function is said to be of class $C^k$ if its derivatives, up to the order $k$, exist and are continuous. For example, the class $C^0$ consists of all continuous functions, while the class $C^1$ consists of all differentiable functions whose derivative exists and is of class $C^0$. In general, for a class $C^0$ function,
\begin{equation} \label{f'}
\int f'(x) dx \ne f(x)~,
\end{equation}
if the integration area contains a point of discontinuity of $f'(x)$, some kind of regularization procedure is needed.

The $C^2$-differentiability assumption plays a key role in the singularity theorems in general relativity (see the reviews \cite{Man-San, Clarke, Gar-San, Chru, Sor-Woo}). Indeed, the standard causality theory assumes that the metric is smooth, or is at least $C^2$, i.e. all the second order partial derivatives of the metric tensor exist and are continuous. Some authors, nevertheless, assume that the components of the metric tensor can be of class $C^1$ (i.e. the second derivatives of the metric tensor, and thus the Riemann tensor, suffer a jump discontinuity), since there are many physically realistic systems of that type, such as the Oppenheimer-Snyder model of a collapsing star \cite{Opp-Sny} and general matched spacetimes, see e.g. \cite{Lich, Synge}. The issue of regularity in the singularity theorems is often ignored despite its mathematical and physical relevance \cite{Sano}.

For the case of a BH, if its exterior region is dynamically stable - as is widely expected - it is inappropriate to possess the Cauchy horizon inside of the BH, since it follows that the $C^0$-inextendibility formulation of the strong cosmic censorship conjecture is false \cite{Dafermos, Klain}. The $C^0$-metric Cauchy horizons are generically singular in an essential way, representing so-called 'weak null singularities'.

We want to emphasize that analysis of the initial value problem for the Einstein equations leads to the restriction that admissible coordinate transformations be of class $C^2$ \cite{Lich, Synge}, they should not change the Riemann tensor, otherwise will lead to the fictitious extra source in the Einstein equations. For example, the class $C^0$ coordinate transformations, $x_\nu \to x_\nu^*$, drop the differentiability of the transformed metric tensor,
\begin{equation} \label{g*}
g_{\mu\nu}^* = g_{\alpha\beta} \frac {\partial x^\alpha}{\partial x^{*\mu}}\frac {\partial x^\beta}{\partial x^{*\nu}}~,
\end{equation}
to $C^0$, even if the initial metric tensor $g_{\alpha\beta}$ was smooth. Then $g_{\mu\nu}^*$ will introduce the $\delta$-like parts in Riemann tensor, i.e. a singular hyper-surfaces in spacetime. This means that two metric tensors, $g_{\alpha\beta}$ and $g_{\mu\nu}^*$, are the solutions of different Einstein equations and do not coincide on the surface of discontinuity. For instance, the spherically symmetric vacuum solutions of Einstein's equations is not diffeomorphic to the metric corresponding to the gravitational field of a $\delta$-like source at the origin \cite{Castro}.

So, in our opinion, the conclusion of absence of physical singularities at a horizon (in the points where all geometrical invariants of the Riemann spacetime are regular), is mathematically correct only for at least class $C^2$ metric tensors. For the lower classes, in addition to the geodesic equations, should be considered the behaviors of quantum particles that obey the equations with second derivatives, with proper boundary conditions at singular points.

The example of misunderstandings with singular coordinate transformations and using of improper boundary conditions is the appearance of the fictitious delta function in the Laplace equation in spherical coordinates \cite{Khel-Nad-1, Khel-Nad-2, Can-Khe}. It is known that the regular solutions to the Laplace equation in Cartesian coordinates,
\begin{equation} \label{Laplace}
\left( \partial_x^2 + \partial_y^2 + \partial_z^2 \right){\cal \varphi} = 0 ~,
\end{equation}
are given by exponential functions of the form
\begin{equation}
{\cal \varphi } \sim e^{\pm [i (ax + by) + \sqrt {a^2 + b^2}z]}~,
\end{equation}
where $a$ and $b$ are integration constants \cite{Jackson}. This solution is regular everywhere and at the origin is constant. After rewriting the Laplace equation (\ref{Laplace}) in spherical coordinates, without setting the adequate boundary conditions at the origin, one can obtain the singular solutions in the form of the Newtonian potential,
\begin{equation}
{\cal \varphi } \sim \frac 1r ~,
\end{equation}
which corresponds to some fictitious $\delta$-source at the origin. So, this solution does not follow from the regular boundary conditions at the origin.

Another well-known example is the two-dimensional potential problem:
\begin{equation} \label{R(C=0)}
R'' + \frac 1r R' = \frac 1r \left(r R'\right)' = 0~,
\end{equation}
where primes denote derivatives with respect to the 2-dimensional radial coordinate, $r = \sqrt{x^2+y^2}$. The famous logarithmic solution to this equation,
\begin{equation} \label{ln}
R \sim \ln \left| \frac 1r \right|~,
\end{equation}
corresponds to a point-like source, $\delta (r)$, and does not represent the solution to (\ref{R(C=0)}) without delta function at the right side, similar to the case with the Laplace case mentioned above. Thus, the regular solution of (\ref{R(C=0)}) at $r = 0$ is not given by (\ref{ln}), but has the form:
\begin{equation}
R(r = 0) = const ~.
\end{equation}


\section{Vacuum Gauge Fields} \label{AB}

The standard assumption from classical physics is that physical effects can be entirely explained by interactions with local field. But there are some quantum mechanical phenomenon that appears to conflict with this notion for the class of singular vector potentials for which the field strength is zero (Vacuum Gauge Fields) \cite{Pesh-Ton},
\begin{equation} \label{F=0}
F_{\mu\nu} \equiv \partial_\mu A_\nu - \partial_\nu A_\mu = 0~.
\end{equation}
Besides the trivial gauges,
\begin{equation}
A_\mu(x) = \partial_\mu \phi(x)~,
\end{equation}
the equation (\ref{F=0}) gives so called topological solutions, which leads to $\delta$-like fields. The well know example of such solution is the Aharonov-Bohm vector potential, which in Cartesian coordinates can be written in the form:
\begin{equation} \label{A-AB}
A_t = A_z = 0~, \qquad A_x = \frac {\alpha y}{x^2 + y^2}~, \qquad A_y = -\frac {\alpha x}{x^2 + y^2}~,
\end{equation}
where $\alpha$ is a non-integer constant. It is obvious that in regular functions, for such potentials, all field invariants ($F_{\mu\nu}F^{\mu\nu}$ and $\epsilon^{\mu\nu\alpha\beta}F_{\mu\nu}F_{\alpha\beta}$) are zero. However, the Aharonov-Bohm effect is that the motion of an electrically charged particle outside the singular region (i.e. fare from $x=y=0$) can be shown to be dependent on (\ref{A-AB}). At the same time, it is known that the classical equations of motion of charged particles with the topologically non-trivial potentials like (\ref{A-AB}) is useless in the singularity region ($x, y \to 0$), where field is $\delta$-like.

Opposite approach is usually used for the free Einstein equations,
\begin{equation} \label{Riemann=0}
R_{\mu\nu} = 0~,
\end{equation}
which looks similar to the condition (\ref{F=0}) and also gives singular solutions, like corresponding to BHs. All invariants (Kretschmann, Chern-Pontryagin and Euler scalars) outside the region of so-called naked singularities are regular. However, besides the trivial coordinate transformations (\ref{g*}) the Vacuum Gravitational Field condition (\ref{Riemann=0}) has topological solutions -- $C^0$-class metric tensors, for which
\begin{equation}
\partial^\mu R_{\mu\nu\alpha\beta} \ne 0 ~.
\end{equation}


\section{Singular transformations of the Schwarzschild metric} \label{Sin-Sch}

As an example of singular spacetimes let us consider the spherically symmetric solution of the vacuum Einstein equations, which in the gauge ascribed to Schwarzschild has the form:
\begin{equation} \label{Schwarzschild}
ds^2 = f(r) dt^2 - \frac {dr^2}{f(r)} - r^2 (d\theta^2 + \sin^2 \theta d\phi^2)~.
\end{equation}
Here we have introduced the notation:
\begin{equation} \label{f}
f(r) = 1 - \frac {r_s}{r}~, \qquad \left \{
\begin{array} {lr}
r_s \leq r \leq \infty\\
0 \leq f \leq 1
\end{array}
\right.
\end{equation}
where the parameter $r_s = 2MG$ determines the Schwarzschild horizon.

To explore the horizon region, the standard approach uses the classical geodesic equations (\ref{geodesic}) and introduces so called Regge-Wheeler's tortoise coordinate \cite{BH-1, BH-2, BH-3},
\begin{equation} \label{tortoise}
r^* = \int \frac {dr}{1-r_s/r} = r + r_s \ln \left( \frac {r}{r_s} - 1\right)~. \qquad \left \{
\begin{array} {lr}
r_s \leq r \leq \infty\\
-\infty \leq r^* \leq \infty
\end{array}
\right.
\end{equation}
Then in the geodesic equations the Schwarzschild horizon 'disappears' in various singular coordinates and cannot prevent classical particles to reach the central naked singularity. However, the coordinate (\ref{tortoise}) is the function of the type (\ref{f'}), i.e. not belongs to the $C^2$-class of admissible coordinate transformations, mentioned in the Sec.~\ref{Sing-Coord}. The singular coordinates used in standard approaches, like introduced by Kruskal-Szekeres, Eddington-Finkelstein, Lema\^{\i}tre, or Gullstrand-Painlev\'{e}, give $\delta$-functions in the second derivatives, since they contain one of the factors
\begin{equation}
\sqrt{r_s - r}~, \quad {\rm or} \quad \ln |r_s - r|~.
\end{equation}
This means that transformed metric tensors at $r=r_s$ are not differentiable, i.e. are of unacceptable class $C^0$ (not of $C^2$, or $C^1$) and the Einstein equation for these metrics is altered with fictitious $\delta$-sources at $r=r_s$. So, while the singular coordinate transformations (which are necessary to hide the horizon singularity) do not cause problems on the level of the geodesic equations, they lead to the appearance of $\delta$-functions at $r=r_s$ in the second order differential equations.

In the Schwarzschild's gauge (\ref{Schwarzschild}), the point-like source of gravity is located at $r = 0$ and its surface area is $4\pi r^2 = 0$.  Also the Kretschmann invariant for (\ref{Schwarzschild}),
\begin{equation} \label{Kretschmann}
R^{\alpha\beta\gamma\delta}R_{\alpha\beta\gamma\delta} = \frac {12 r_s^2}{r^6} ~,
\end{equation}
contains in denominator the Schwarzschild coordinate $r$ (rather than any general function) and blows up only at $r=0$. This tells that the source of gravity is at $r = 0$ and this is again a strong argument that only the Schwarzschild gauge gives a physically meaningful picture. For instance, applying the principle of invariance of 4-volume for Schwarzschild and Eddington-Finkelstein coordinate systems, where the determinant and space-like coordinate variables are the same, one obtains $dt=dt^*$. But in Eddington-Finkelstein coordinates
\begin{equation}
dt^* = dt + \frac {r_s}{r-r_s}~,
\end{equation}
so its follows that mass of the source $M \sim r_s$ should be zero \cite{ADM, Mitra-1, Mitra-2}. For an extended distribution of mass, one must consider suitable static spherically symmetric interior solutions and only for the Schwarzschild gauge the gravitational mass of the body is given by
\begin{equation}
M = \int 4\pi \rho (r) r^2 dr~.
\end{equation}
To the best of the knowledge, no corresponding study has ever been made for other choices of gauge and there is no known formula for $M$ in other coordinates. For other gauges, one can arrive to some paradoxes, e.g. obtain the point mass with finite surface are.

Now let as concern on the horizon region of (\ref{Schwarzschild}), where the Kretschmann scalar  (\ref{Kretschmann}) is seems to be regular. For sufficiently large values of BH mass $M$ one can let (\ref{Kretschmann}) to be arbitrarily small at $r = r_s = 2MG$. From the other hand we know that the gravity at the horizon is supposed to be so strong that even light cannot escape it. But how the weak gravity trap light? Obviously, this would be unphysical. Thus conclusion on the strength of gravity close to a horizon should be made from the second order equations of test particles that contain derivatives of the Riemann tensor, as it was mentioned in Sec.~\ref{AB}. While the expression (\ref{Kretschmann}) at $r = r_s$ is obtained from the assumption of the type $0/0 = 1$. The same is true for the determinant of Schwarzschild's metric tensor, where the product of its components, $g_{tt}\cdot g_{rr}$, is ill-defined at $r = r_s$. In general, $g_{tt}$ and $g_{rr}$ are independent functions and the cancelation of their zeros is accidental, since follows from the validity of the vacuum Einstein equations. However, exact spherical symmetry and true vacuums are rarely, if ever, observed.

For instance, already on the classical level, the second derivatives of the metric tensor (i.e. the Riemann tensor) enters the equations of motion of a system of particles in the quadruple approximation \cite{Dix},
\begin{equation} \label{quad}
\frac{Dp^\mu}{ds} = F^\mu = - \frac 12 R^\mu{}_{\nu \alpha\beta}u^\nu S^{\alpha \beta} - \frac 16 J^{\alpha\beta\gamma\delta}D^\mu R_{\alpha\beta\gamma\delta}~,
\end{equation}
where $J^{\alpha\beta\gamma\delta}$ is the quadruple moment of the source, $S^{\alpha \beta}$ is the spin tensor and $u^\nu$ is the 4-velocity. Therefore, the force, $F^\mu$, diverges at the Schwarzschild horizon, since the three from six non-zero independent components of the mixed Riemann tensor,
\begin{equation}
R^t{}_{rrt} = 2R^\theta{}_{r\theta r} = 2R^\phi{}_{r\phi r} = \frac {r_s}{r^2(r_s - r)}~,
\end{equation}
blow up at $r = r_s$.


\section{Klein-Gordon equation vs Hamilton-Jacobi equation} \label{KG-HJ}

We note that, like the Aharonov-Bohm case considered in Sec.~\ref{AB}, non-trivial effects for singular potentials is expected to appear for quantum particles. So, to explore singularity regions it is not enough to study only classical geodesic equations.

It is known that classical trajectories of particles can be obtained from the wavefunctions of quantum particles in geometrical-optical limit (eikonal approximation) \cite{Gold}. Einstein's gravity doesn't care about spin and to explore strong gravitational fields one can restrict himself with the Klein-Gordon equation,
\begin{equation} \label{wave}
(\Box + \mu^2)\Phi = \left[\frac {1}{ \sqrt{-g}}\partial_\mu \left( \sqrt{-g}g^{\mu\nu}\partial_\nu \right) + \mu^2 \right]\Phi = 0~.
\end{equation}
For photons the mass parameter is zero, $\mu = 0$, while for fermions one needs to obtain the single equation of the type (\ref{wave}) from the first order Dirac's system.

The Klein-Gordon wave functions $\Phi$ associated with the classical motion formally obey the relativistic Hamilton-Jacobi equation \cite{Motz},
\begin{equation} \label{geodesic}
g_{\mu\nu}p^\mu p^\nu - \mu^2 = 0~,
\end{equation}
where $p^\nu = \mu ~dx^\nu/ds$ denotes relativistic 4-momentum that contain the first derivatives of coordinate functions. Indeed, in the semi-classical approximation the scalar wave function in (\ref{wave}) can be expressed in terms of an amplitude and phase,
\begin{equation} \label{Phi=rho}
\Phi = A e^{iS}~,
\end{equation}
where $S(x^\nu)$ is the Hamilton principal function, which usually is used in the definition of the classical momentum,
\begin{equation} \label{S'}
p^\nu \sim \partial^\nu S ~.
\end{equation}
Then the Klein-Gordon equation (\ref{wave}) gives the system of equations:
\begin{equation} \label{KG-system}
\begin{split}
A \Box S + 2 \partial_\nu S \partial^\nu A = 0~, \\
\Box A - A \partial_\nu S \partial^\nu S + \mu^2 A = 0~.
\end{split}
\end{equation}

The approximations needed to reduce the Klein-Gordon system (\ref{KG-system}) to the Hamilton-Jacobi (geodesic) equation (\ref{geodesic}) are:
\begin{itemize}
\item (i) Week gravitational field and short wavelength, $\Box S \to 0$;
\item (ii) Negligible variations of the wave amplitude, $\partial^\nu A \to 0$.
\end{itemize}
From the condition (i) it follows that the eikonal phase (Hamilton's principal function) can be written as:
\begin{equation} \label{S}
S \sim p_\nu x^\nu~,
\end{equation}
where $p_\nu$ obeys (\ref{geodesic}). Close to a singularity the approximations (i) and (ii) are not valid and to explore this region one needs to consider the full Klein-Gordon system (\ref{KG-system}), not only the geodesic equations (\ref{geodesic}). Unlike the Hamilton-Jacobi equation (\ref{geodesic}), the Klein-Gordon equation (\ref{wave}) contains second derivatives of the particle wavefunction and without inserting of extra sources does not give the class $C^0$ physical solutions.


\section{Klein-Gordon equation in Schwarzschild spacetime} \label{KG-Sch}

Consider the infalling particles, which obey the Klein-Gordon equation (\ref{wave}), on s Schwarzschild BH. The Schwarzschild metric (\ref{Schwarzschild}) is highly symmetric and we can separate the variables,
\begin{equation} \label{Phi}
\Phi \sim \psi (t, r) Y_{lm}(\theta, \phi) ~,
\end{equation}
where $Y_{lm}(\theta, \phi)$ are spherical harmonics. Then the Klein-Gordon equation (\ref{wave}) gives:
\begin{equation} \label{psi}
\left\{ r^2 \partial_t^2 - f \partial_r\left( r^2 f\partial_r\right) + f\left[l(l+1) + r^2 \mu^2 \right]\right\}\psi (t,r) = 0~,
\end{equation}
where $l$ is the orbital angular momentum quantum number. Close to the horizon, $f \to 0$, the mass and angular momentum terms in (\ref{psi}) can be ignored,
\begin{equation} \label{m,l=0}
\mu, l \to 0~.
\end{equation}
Then after the additional separation of variables,
\begin{equation} \label{psi=RT}
\psi (t,r) = \frac 1r T(t)R(r) ~,
\end{equation}
the equation (\ref{psi}) gives the system:
\begin{equation} \label{T''}
\partial^2_t T = C T ~,
\end{equation}
\begin{equation} \label{R''}
f^2 \partial_r^2 R + \frac{r_sf}{r^2} \partial_r R - \left(C + \frac {r_sf}{r^3}\right) R = 0~,
\end{equation}
where $C$ is a separation constant.

To study behaviors of quantum particles close to the horizon one needs to obtain the solutions to the system (\ref{T''}) - (\ref{R''}) with the adequate boundary conditions at $f = 0$. In standard approach, the boundary conditions for (\ref{T''}) - (\ref{R''}) is usually settled assuming the existence of a horizon crossing, ingoing radial waves\cite{Star,Matz},
\begin{equation} \label{solution-f=0}
T(t)_{f \to 0} \sim  e^{\pm i\omega t}~, \qquad R(r)_{f \to 0} \sim e^{\pm i\omega r^*} ~,
\end{equation}
where the Regge-Wheeler tortoise coordinate (\ref{tortoise}) is introduced. This brings (\ref{R''}) to a Schr\"{o}dinger-like equation,
\begin{equation} \label{KG-R}
\frac {d^2R}{dr^{*2}} - \left[V(r^*) + C\right] R = 0 ~,
\end{equation}
where the effective potential for the simplest case (\ref{m,l=0}) is:
\begin{equation} \label{Veff}
V(r^*) = \frac {r_sf(r^*)}{r^3(r^*)} ~.
\end{equation}
The assumption (\ref{solution-f=0}) corresponds to the negative separation constant in the system (\ref{T''}) - (\ref{R''}),
\begin{equation}
C = -\omega^2 < 0 ~.
\end{equation}
However, as it was mentioned above, the transformation (\ref{tortoise}) is singular and the solution (\ref{solution-f=0}) does not obey the equation (\ref{R''}), due to the appearance of the $\delta$-function in the second derivatives of the tortoise coordinate function (\ref{tortoise}) at the horizon, $r=r_s$.

To clarify this point let us consider the model equation,
\begin{equation} \label{cal-R}
r(r-1)^2 \partial_r^2 {\cal R} + (r-1) \partial_r {\cal R} + r^3 {\cal R} = 0~,
\end{equation}
which has similar to (\ref{R''}) kinetic part and for simplicity it is assumed that $r_s = 1$. Due to the third term in this equation, it is obvious that a regular solution to (\ref{cal-R}) at $r\to 1$, should obey the boundary condition:
\begin{equation} \label{}
{\cal R}|_{r\to 1} \to 0~.
\end{equation}
From the other hand, using the singular tortoise coordinate (\ref{tortoise}) (with $r_s = 1$),
\begin{equation} \label{y}
y = r + \ln \left( r - 1\right)~,
\end{equation}
the equation (\ref{cal-R}) can be transformed to the form:
\begin{equation} \label{cal-r}
\partial_y^2 {\cal R} + {\cal R} = 0~.
\end{equation}
Without considering the existence of a $\delta$-function in (\ref{cal-r}), one can obtain the similar to (\ref{solution-f=0}) non-physical harmonic solution,
\begin{equation}
{\cal R} \sim e^{\pm iy} ~,
\end{equation}
which is valid in the whole space of the new variable, $-\infty \leq y \leq \infty$, including the horizon $y \to -\infty$, i.e. when $r \to 1$.

To find correct boundary conditions at the BH horizon let us use $f(r)$ as an independent variable in the region $r_s \le r \le \infty$ ($0 \le f \le 1$) in (\ref{R''}) and rewrite this equation in the form:
\begin{equation} \label{R(f)}
f^2(1-f)^4 R'' + f(1-f)^3(1-3f) R' -\left[r_s^2C + f(1-f)^3 \right] R = 0~,
\end{equation}
where primes denote derivatives with respect to $f$. Close to the horizon this equation takes the asymptotic form:
\begin{equation} \label{R(f=0)}
f^2R'' + f R' - r_s^2C R = 0~. \qquad (f \to 0)
\end{equation}
To set the physical boundary condition at the horizon we need to find a regular solution to this equation at $f = 0$. Note that the case with zero separation constant in (\ref{R(f=0)}), $C=0$, exactly coincides with the 2-dimensional potential equation (\ref{R(C=0)}), considered in the Sec. \ref{Sing-Coord}.

Using these observations we see that the equation (\ref{R(f=0)}) has a regular at $f=0$ solution only for a positive separation constant $C$ \cite{Qin, Gog-Pan},
\begin{equation} \label{R->0}
R(r)|_{f \to 0} \sim f^{r_s\sqrt C} ~. \qquad (C > 0)
\end{equation}
Again we note that the solution to (\ref{R(f=0)}) with the negative $C$,
\begin{equation} \label{C<0}
R(r)|_{f \to 0} \sim e^{i r_s\sqrt {-C} \ln |f|} ~, \qquad (C < 0)
\end{equation}
which is used as a boundary solution in (\ref{solution-f=0}), does not obey (\ref{R''}), since it leads to the extra delta-like source term at the BH horizon.

From the boundary solution (\ref{R->0}), which we want to use below, follows that
\begin{equation} \label{R=0}
R(r)_{f \to 0} \rightarrow 0 ~,
\end{equation}
i.e. the probability for particles to cross the horizon is tends to zero and the infalling particles are absorbed by a BH or can be reflected from its horizon \cite{Gog-Pan}. In this case the information about quantum particles does not disappear for a distance observer, which can solve the BH information paradox \cite{Mar}.

Now let us look for the solution to (\ref{R(f)}) for the regular boundary condition (\ref{R->0}), using the method of Frobenius:
\begin{equation} \label{R=}
R (f) = \sum_{i}^{\infty} a_{i} f^{i}~, \qquad (i = 1, ...,\infty)
\end{equation}
where $a_{i}$ are some constants. The equation (\ref{R(f)}) reduces to the algebraic system for the coefficients of $f^i$, which will be satisfied when the coefficients of each $f^i$ becomes zero. The first equation of this system ($i = 1$),
\begin{equation} \label{a}
f(1-f)^3(1-3f)a_1 -\left[r_s^2C + f(1-f)^3 \right]a_1f \approx \left(1 - r_s^2 C \right)\left(a_1f\right) = 0 ~,
\end{equation}
shows that the separation constant in the system (\ref{T''}) - (\ref{R''}) indeed is positive \cite{Qin},
\begin{equation} \label{k}
C \approx \frac {1}{r_s^2} > 0 ~.
\end{equation}
In this case from (\ref{T''}) we find the real-valued exponentially time-dependent solution (with the complex phase $\omega = \pm i/r_s$),
\begin{equation} \label{T}
T(t)= T_0 e^{\pm t/r_s} ~,
\end{equation}
i.e. in the Schwarzschild coordinates of a distant observer the wave function (\ref{psi=RT}) has the form \cite{Gog-Pan}:
\begin{equation} \label{psi=}
\psi(t,r) \sim \frac {e^{\pm t/r_s}}{r} \sum_{i}^{\infty} a_{i} \left(1 - \frac {r_s}{r}\right)^{i} ~.
\end{equation}
The constants $a_i$ obey the conditions,
\begin{equation}
\frac{a_{i+1}}{a_i}\big|_{i \to \infty} \to 2~,
\end{equation}
what means that the condition of convergence of the radial wave function (\ref{R=}),
\begin{equation}
\frac{a_{i+1} f^{i+1}}{a_i f^i}\big|_{i \to \infty} < 1 ~,
\end{equation}
is satisfied for
\begin{equation} \label{f<1/2}
0 < f < \frac 12~. \qquad (r_s <r < 2r_s)
\end{equation}
Thus the solution (\ref{psi=}) is valid in the interesting for us region close to the Schwarzschild sphere.

The obtained solution with the complex phase (\ref{psi=}) is very different from the familiar internal \cite{BH-wave-1, BH-wave-2} and external \cite{BH-out} periodic-in-time solutions ($\sim e^{\pm i\omega t}$) for the scalar particles in Schwarzschild. In semi-classical quantum mechanics, a wave function with the complex phase usually is the signature of a tunneling processes through a potential barrier, i.e. the penetration of particles through the BH horizon cannot occur classically. So, our exponentially decreasing (increasing) in time solutions (\ref{T}) show that quantum particles probably do not cross the Schwarzschild horizon at all, but are absorbed or reflected by it. This potentially solves the BH singularity problems, since the event horizon appears to be the edge of an impenetrable sphere with ultra-dense matter inside, where nothing disappears and quantum effects smooth out central singularity.

Note that the solutions to the Klein-Gordon equation in Schwarzschild's coordinates having the complex phase was obtained also in \cite{Tunneling-1, Tunneling-2, Tunneling-3, Tunneling-4}, were it was nevertheless assumed that classical geodesics are extendable across the horizon, while the real-valued exponential factor was connected with the process of particle creation by the gravitational field of BHs. But in these papers the singular point at the BH horizon, which shows that classical particles are stopped from entering the Schwarzschild sphere, was removed by the introduction of the infinitesimal integration contours around the pole in the propagator of scalar field.


\section{Klein-Gordon equation in Kerr spacetime} \label{KG-Kerr}

Realistic BHs are formed from the collapsing stars and should have an angular momentum. Therefore, it is important to demonstrate that quantum particles do not cross the horizon for Kerr's BH as well, what excludes the BH singularity problem.

Let us consider a quantum particle close to the Kerr BH event horizon and demonstrate that there is a damping of the wave function, like in the Schwarzschild case (\ref{psi=}). The Klein-Gordon equation in the field of uncharged Kerr's BH of the mass $M$ and with the angular parameter $a$ can be written in the form (see, for example \cite{Vi-Be-Mu}):
\begin{equation}\label{eq:6}
\begin{split}
\left\{ \frac{1}{\Delta} \left[ (r^2 + a^2)^2 - \Delta a^2 sin^2 \theta \right] \frac{\partial^2}{\partial t^2}
- \frac{\partial}{\partial r} \left( \Delta \frac{\partial}{\partial r} \right)
- \frac{1}{sin\theta} \frac{\partial}{\partial \theta} \left( sin\theta \frac{\partial}{\partial \theta} \right) - \right.\\
\left. - \frac{1}{sin^2\theta} \left( \Delta - a^2 sin^2\theta \right) \frac{\partial^2}{\partial \phi^2}
+ \frac{2a}{\Delta} \left( r^2 + a^2 - \Delta \right) \frac{\partial^2}{\partial t \partial \phi} + \mu^2\right\} \Psi = 0 ~,
\end{split}
\end{equation}
where $\Delta = r^2 - 2Mr + a^2$. Let us suppose that in the Kerr spacetime it is possible to separate variables in the wavefunction in the form:
\begin{equation} \label{eq:7}
\Psi = e^{-\omega t} R(r)S(\theta)e^{im\phi} ~.
\end{equation}
Unlike the common case, here we have chosen the real-valued time dependence, $e^{-\omega t}$, since in the limit $a \rightarrow 0$ the solution of (\ref{eq:6}) should reduce the Schwarzschild one (\ref{psi=}). This leads to the additional restriction, the mixed term in (\ref{eq:6}) (which contains $\partial^2/\partial t \partial \phi$ and is absent in the Schwarzschild case) becomes complex and to eliminate it we need to set the azimuthal quantum number to zero,
\begin{equation} \label{m=0}
m = 0~.
\end{equation}

Substitution of (\ref{eq:7}) into (\ref{eq:6}) gives the following system of equations:
\begin{eqnarray} \label{eq:8}
\left(\frac{d^2}{d\theta^2} + \frac{cos\theta}{sin\theta} \frac{d}{d \theta} + a^2 \omega^2 sin\theta  - \frac{m^2}{sin^2 \theta} - C \right) S (\theta) = 0 ~,\\
\label{eq:9}
\left[\frac{d^2}{dr^2}+ \frac{2(r - M)}{\Delta}\frac{d}{dr} + \frac{m^2a^2}{\Delta^2} + \frac{i 2am\omega}{\Delta^2}\left(r^2 + a^2 - \Delta\right)  - \frac{\omega^2}{\Delta^2}\left(r^2 + a^2\right) + \frac {C - \mu^2}{\Delta} \right] R(r) = 0 ~,
\end{eqnarray}
where $C$ is the separation constant.

Changing in (\ref{eq:9}) the radial variable, $r$, by the physical variable, $\Delta$, and setting (\ref{m=0}) (to make this equation real), we find,
\begin{equation} \label{eq:10}
\begin{split}
4\Delta^2\left(M^2 - a^2 + \Delta\right) \frac{d^2 R}{d\Delta^2} + \Delta \left[ 4\left(M^2 - a^2 + \Delta\right) + 2\Delta \right] \frac{dR}{d \Delta} + \\
+ \left[ \left(C - \mu^2\right)\Delta - \omega^2 \Delta^2 - 2M \omega^2 \left(M \pm \sqrt{M^2 - a^2 + \Delta} \right)\right] R (\Delta) = 0 ~.
\end{split}
\end{equation}
Near the horizon, $\Delta \rightarrow 0$, this equation takes the form:
\begin{equation} \label{eq:13}
\left(\Delta^2 \frac{d^2}{d\Delta^2} + \Delta \frac{d}{d\Delta} - \kappa^2 \right) R (\Delta) = 0~,
\end{equation}
where we had introduced the parameter,
\begin{equation}
\kappa^2 = \frac {M \omega^2 \left( M \pm \sqrt{M^2 - a^2}\right)}{2 (M^2 - a^2)} ~.
\end{equation}
The regular solution to the equation (\ref{eq:13}),
\begin{equation} \label{eq:14}
R \propto \Delta^\kappa ~,
\end{equation}
indicates that as a particle moves toward the horizon, i.e. $\Delta \rightarrow 0 $, its radial wavefunction tends to zero,
\begin{equation}
R (\Delta) \rightarrow 0 ~,
\end{equation}
as for the Schwarzschild case (\ref{R=0}), i.e. quantum particles do not cross the Kerr BH horizon as well.

For completeness let us show that another component of the wavefunction (\ref{eq:7}), $S(\theta)$, stays finite at the Kerr horizon, $\Delta = 0$. The change of the variable, $\cos\theta \equiv x$, brings the equation (\ref{eq:8}) to the form:
\begin{equation} \label{eq:17}
\left(1 - 2x^2 + x^4\right)\frac{d^2 S}{d x^2} - 2\left(x - x^3\right)\frac{d S}{d x} - \left(m^2 + 1 + x^2\right)S = 0~,
\end{equation}
which has the solution in Legendre polynomials, i.e. $S(\theta)$ is a finite function.

So, as for the Schwarzschild case, quantum particles do not enter the Kerr BHs horizon, but are reflected or absorbed by it. The new result here is that, due to the coupling of the particles and Kerr's BH angular momentums, this reflection or absorbtion is possible only when the infalling particle loses its angular momentum, i.e. for the case of zero azimuthal quantum number, $m = 0$.


\section{Conclusion} \label{Concl}

In this article we offer the view that a Schwarzschild or Kerr BH horizon might represent more singular surface than is usually thought and represents a physical barrier to particles/fields trying to cross it. This is in contrast to the standard view that the horizon is simply a coordinate singularity. To demonstrate this we note that the Regge-Wheeler's 'tortoise' radius and the singular coordinate transformations of metric tensor, used to demonstrate continuity of geodesics at a BH horizon, are class $C^0$ functions. However, the standard causality theory requires that the metric tensor to be at least $C^1$ function (whose derivative exists and is continuous). For the case when components of the transformed metric tensor are class $C^0$, its first derivatives are discontinues and the second derivatives lead to the appearance of the fictitious $\delta$-like sources in the Einstein equations. So, while the singular coordinate transformations (which are necessary to hide the horizon singularities) do not cause problems on the level of the classical geodesic equations (which contain the first derivatives of the coordinate functions), they lead to the appearance of $\delta$-functions in the equations of quantum particles (which contain the second derivatives of wavefunctions). Therefore, to explore the horizon region of the black hole, one needs to use solutions of the exact second order equation of particles and not only the classical geodesic equations.

To justify this claim we solve the Klein-Gordon equation on the Schwarzschild and Kerr backgrounds with the physical boundary conditions at the event horizons (obtained without performing singular coordinate transformations). We have found the real-valued exponentially time-dependent solutions (with the complex phases), i.e. in semi-classical approximation particles probably do not enter the horizon around of a compact object, but are absorbed and some are reflected by it. For the Kerr case we have come up with the additional criteria for the particle to be reflected or absorbed, which is that the azimuthal quantum number should be zero. Note that for quantum particles still there is a small but finite probability to tunnel through the delta-function barrier at the horizon.

One consequence of our observations is that minimal radius of any isolated classical body is its event radius, inside of which classical General Relativity is not valid. This potentially solves the BH singularity and information problems, since the event horizon seems to surround an impenetrable sphere of the ultra-dense matter where quantum effects smooth out central singularity (similar to the case with a point-like source in classical physics) and also physical information couldn't disappear under the event horizon. Moreover, we show that the reflected from the horizon particles could obtain energy from the gravitational field and these glints can explain some burst-type signals in the universe, like GRBs, FRBs, ultra-high energy cosmic rays, or the LIGO events \cite{Gog-Pan}.


\section*{Acknowledgments}

This work was supported by Shota Rustaveli National Science Foundation of Georgia (SRNSFG) [DI-18-335/New Theoretical Models for Dark Matter Exploration].


\end{document}